\documentclass{aa}
\newcommand{\CI}{[\ion{C}{i}]}
\usepackage{graphicx}
\usepackage{txfonts}
\usepackage{natbib}
\usepackage{color}
\begin{document}
   \title{The Photon Dominated Region in the IC\,348 molecular cloud}

   \author{K. Sun\inst{1}, V. Ossenkopf\inst{1,2}, C. Kramer\inst{1}, B. Mookerjea\inst{3}, M. R\"ollig\inst{4}, M. Cubick\inst{1} \and J. Stutzki\inst{1}}

   \offprints{K. Sun,\\ \email{kefeng@ph1.uni-koeln.de}}

   \institute{I. Physikalisches Institut, Universit\"at zu K\"oln, Z\"ulpicher Stra\ss e 77, D-50937 K\"oln, Germany
         \and
             SRON Netherlands Institut for Space Research, P. O. Box 800, 9700, AV Groningen, the Netherlands
         \and
             Department of Astronomy \& Astrophysics, Tata Institute of Fundamental Research, Homi Bhabha Road, Mumbai 400 005, India
         \and
             Argelander-Institut f\"ur Astronomie, Universit\"at Bonn, Auf dem H\"ugel 71, D-53121 Bonn, Germany
             }

   \date{Received; accepted}
   \authorrunning{Sun et al.}
   \titlerunning{The Photon Dominated Region in the IC\,348 molecular cloud}
% \abstract{}{}{}{}{}
% 5 {} token are mandatory

  \abstract
   {}
   {In this paper we discuss the physical conditions of clumpy nature in the
   IC\,348 molecular cloud. We show that the millimeter and sub-millimeter line emission from
the IC\,348 molecular cloud
   can be modelled as originating from a photon dominated region (PDR).}
   {We combine new observations of fully sampled maps in [\ion{C}{i}] at 492\,GHz and $^{12}$CO 4--3,
    taken with the KOSMA 3~m telescope at about 1\arcmin ~resolution, with FCRAO data of
    $^{12}$CO 1--0, $^{13}$CO 1--0 and far-infrared continuum data observed by HIRES/IRAS.
    To derive the physical parameters of the region we analyze the line ratios of
    [\ion{C}{i}] $^{3}$P$_{1}$--$^{3}$P$_{0}/^{12}$CO 4--3,
    [\ion{C}{i}]$^{3}$P$_{1}$--$^{3}$P$_{0}/^{13}$CO 1--0, and $^{12}$CO 4--3$/^{12}$CO 1--0.
    A first rough estimate of abundance is obtained from an LTE analysis. To understand the
    [\ion{C}{i}] and CO emission from the PDRs in IC\,348, we use a clumpy PDR model. With an
    ensemble of identical clumps, we constrain the total mass
    from the observed absolute intensities. Then we apply a more realistic clump distribution model with a power law index of 1.8
    for clump-mass spectrum and a power law index of 2.3 for mass-size relation.}
   {We provide detailed fits to observations at seven representative positions in the
   cloud, revealing clump densities between 4\,10$^{4}$\,cm$^{-3}$ and
   4\,10$^{5}$\,cm$^{-3}$ and C/CO column density ratios between 0.02 and 0.26.
   The derived FUV flux from the model fit is consistent with the field calculated from FIR
   continuum data, varying between 2 and 100 Draine units across the cloud. We find that
   both an ensemble of identical clumps and an ensemble with a power law clump mass distribution produce line intensities
   which are in good agreement (within a factor $\sim 2$) with
   the observed intensities. The models confirm the anti-correlation between the C/CO abundance ratio and the hydrogen column
   density found in many regions.}
   {}

   \keywords{ISM: clouds -- ISM: dust, extinction -- ISM: structure -- ISM: IC\,348}

   \maketitle
%
%________________________________________________________________

\section{Introduction}

Photon Dominated Regions (PDRs) are surfaces of molecular clouds
where chemistry and heating are regulated by far-ultraviolet (FUV)
photons (6.0 eV $<$ $h\nu$ $<$ 13.6 eV ) \citep{ht99}. Gas cooling
happens mainly via the fine structure lines of [\ion{O}{i}],
[\ion{C}{ii}], [\ion{C}{i}] and the rotational lines of CO
\citep{kwhl99}. The form of carbon changes with increasing depth
from the surface of the PDR from C$^{+}$ through C$^{0}$ to CO.
Therefore emission from [\ion{C}{ii}], [\ion{C}{i}] and the
rotational lines of CO can be used as probes of temperature,
density and column density in the PDRs \citep{kjmsbs04,mkrm06}. In
particular, the two fine structure lines of neutral atomic carbon
at 492 and 809\,GHz give information on intermediate regions
between atomic and molecular gas, since they trace both physical
regimes. Initial homogeneous plane-parallel steady-state models of
PDRs predicted that PDRs are confined to the FUV irradiated
surfaces of molecular clouds. However, observations detected
widespread [\ion{C}{ii}] and [\ion{C}{i}] emission, thus
suggesting that PDR surfaces are formed deeply inside the
molecular clouds. One possible explanation is in terms of the
clumpiness of the molecular clouds which allows for deeper
penetration of FUV radiation so that PDRs are formed at the
surfaces of the clumps present throughout the molecular cloud
\citep{ssghjb88,hjgs91,mkrm06}. In addition, an explanation has
also been proposed in terms of the dynamics of the ISM: a. clouds
are suddenly shielded from FUV fields \citep{sss97}; b.
time-dependent chemistry of the PDRs \citep{oim04}.

IC\,348 is a young open cluster northeast of the Perseus cloud at
a distance of 316\,pc \citep{her98}. The IC\,348 cluster has about
300 stellar members, with spectral types from B5 to late M, as
found from observations in X-rays \citep{pzh96,pz01},
near-infrared \citep{ll95,lrll98,ntc00,llme05}, and optical
\citep{her98,lsmrlbl03}. \citet{lada06} constructed
optical-infrared spectral energy distributions (SEDs) for all
known members of the cluster by combining the Spitzer Space
Telescope and ground-based observations. The IC\,348 molecular
cloud prominently shows up in the CO 3--2 survey of the Perseus
molecular cloud complex \citep{skobsm06}. The FUV radiation of its
youngest member, HD\,281159, a B\,5\,V star, heats the surrounding
gas and dust. Surveys for molecular outflows revealed so far four
bipolar outflows powered by deeply embedded young sources
\citep{efdm03,tkb06}. \citet{jjkm07} estimated the current star
formation efficiency of 10\%--15\% for the whole Perseus cloud
using Spitzer mid-infrared data and SCUBA sub-millimeter maps.

\citet{skobsm06} observed IC\,348 in four low-$J$ CO transitions
as part of their 7.1 square degree survey of the entire Perseus
molecular cloud and quantified the spatial structure of the
emission in terms of its power spectrum, applying the
$\Delta$-variance method \citep{sbhoz98} to the integrated
intensity maps. The $\Delta$-variance spectra of the overall
region was found to follow a power law with an index of $\beta =
2.9-3.0$; while sub-regions of Perseus gave different power
spectral indices. The active star forming regions showed higher
indices typical for large condensations; lower indices were
derived for dark clouds, representing more filamentary structure.
IC\,348 as an exception, showed a rather filamentary structure
corresponding to the smallest value of $\beta = 2.71$ in $^{13}$CO
2--1.

In this paper we present large-scale (20\arcmin $\times$
20\arcmin) fully-sampled maps of the [\ion{C}{i}] $^{3}$P$_{1}$ --
$^{3}$P$_{0}$ (hereafter [\ion{C}{i}]) and $^{12}$CO 4--3 emission
from the IC\,348 molecular cloud. At a common resolution of
70\arcsec, we have combined our data with the $^{12}$CO 1--0,
$^{13}$CO 1--0 data from the Five College Radio Astronomy
Observatory (FCRAO) \citep{rwmam03}, and the far-infrared (FIR)
continuum data from HIRES/IRAS.  The goal of the analysis is to
understand the extent to which the observed line emission from
IC\,348 can be understood in the framework of a photon dominated
region.

Section~\ref{sec_data} describes the KOSMA observations and the
complementary datasets used for analysis. Section~\ref{sec_obs}
discusses the general observational results. A Local Thermodynamic
Equilibrium (LTE) analysis of the observed line intensities is
presented in Section~\ref{sec_lte}. A comparison with clumpy PDR
models is presented in Section~\ref{sec_pdr}.
Section~\ref{sec_sum} summarizes the results.

%________________________________________________________________

\section{Datasets}\label{sec_data}

\subsection{[\ion{C}{i}] and $^{12}$CO 4--3 observations with KOSMA}

We have used the K\"olner Observatorium f\"ur Sub-Millimeter
Astronomie (KOSMA) 3-m sub-millimeter telescope on Gornergrat,
Switzerland \citep{wbe86,kdg98a} to observe the emission of the
fine structure line of neutral carbon at 492\,GHz and $^{12}$CO
4--3 at 461\,GHz. We have used the Sub-Millimeter Array Receiver
for Two frequencies (SMART) on KOSMA for these observations
\citep{ghm02}. SMART is an eight-pixel dual-frequency
SIS-heterodyne receiver capable of observing in the 650 and 350
$\mu$m atmospheric windows \citep{ghm02}. The IF signals were
analyzed with two 4$\times$1\,GHz array-acousto-optical
spectrometers with a spectral resolution of 1.5\,MHz
\citep{hss99}. The typical double side band receiver noise
temperature at 492\,GHz is about 150\,K. The observations were
performed in position-switched On-The-Fly (OTF) between December
2004 and February 2005. Owing to technical difficulties the higher
frequency channel of SMART could not be used at the time of these
observations.

We observed a fully sampled map of the IC\,348 molecular cloud
centered at $\alpha$ = 03$^{h}$44$^{m}$10$^{s}$, $\delta$ =
32\degr06\arcmin (J2000), extending over 20\arcmin $\times$
20\arcmin. For the observations, we used the position-switched
on-the-fly (OTF) mode \citep{bkds00} with a sampling of 29\arcsec.
The emission-free off position was selected from the $^{13}$CO
1--0 FCRAO map, which is (-8\arcmin,10\arcmin) relative to the map
center. We estimate the pointing accuracy to be better than
10\arcsec. The half power beam width (HPBW) and the main beam
efficiency ($B_\mathrm{eff}$) were derived from continuum scans of
the Sun and the Jupiter. The HPBW at both frequencies is 60\arcsec
and the main beam efficiency is 50\%. The forward efficiency
$F_\mathrm{eff}$ of the telescope is $\sim 90$\%. Atmospheric
calibration was done by measuring the atmospheric emission at the
OFF-position and using  a standard atmospheric model to fit the
opacity taking into account the sideband imbalance
\citep{cern85,hiya98}.

All data presented in this paper are in units of main beam
temperature $T_\mathrm{mb}$, calculated from the observed
calibrated antenna temperature $T_\mathrm{A}^\mathrm{*}$ using the
derived beam and forward efficiencies, $T_\mathrm{mb}$ =
$T_\mathrm{A}^\mathrm{*}$ ($F_\mathrm{eff}/B_\mathrm{eff}$). Based
on observations of reference sources such as W~3 and DR~21 we
estimate the accuracy of the  absolute intensity calibration to be
better than 15\%. The data reduction was carried out using the
GILDAS \footnote{http://www.iram.fr/IRAMFR/GILDAS/} software
package.

\subsection{Complementary data sets}

We have used the FCRAO $^{12}$CO 1--0 and $^{13}$CO 1--0 datasets
with an resolution of 46\arcsec presented by \citet{rwmam03} for
comparison. For comparison, we have smoothed all CO and \ion{C}{i}
data to a common resolution of 70\arcsec assuming that both the
FCRAO and KOSMA telescopes have a Gaussian beam.

Further, we have obtained HIRES processed IRAS maps of the dust
continuum emission at 60 and 100~$\mu$m \citep{afm90}. The dust
continuum maps have resolution of 1\farcm5, which is only slightly
worse than that of our submillimeter datasets.

%________________________________________________________________

\section{Observational results}\label{sec_obs}

\begin{figure*} %[h]
\centering \mbox{
\begin{minipage}[b]{6cm}
\includegraphics[width=6cm,angle=-90]{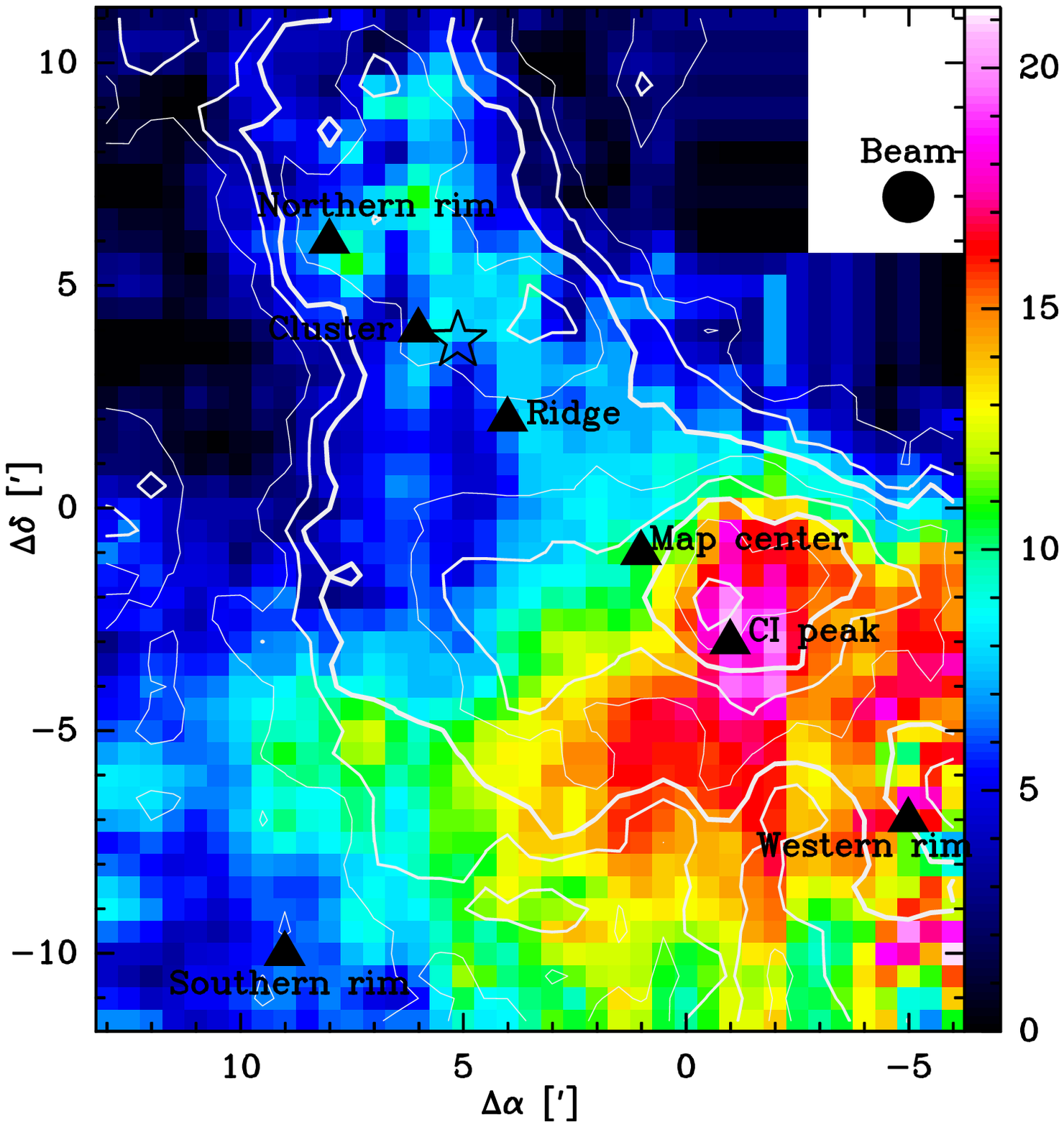}
\end{minipage}
\begin{minipage}[b]{6cm}
\includegraphics[width=6cm,angle=-90]{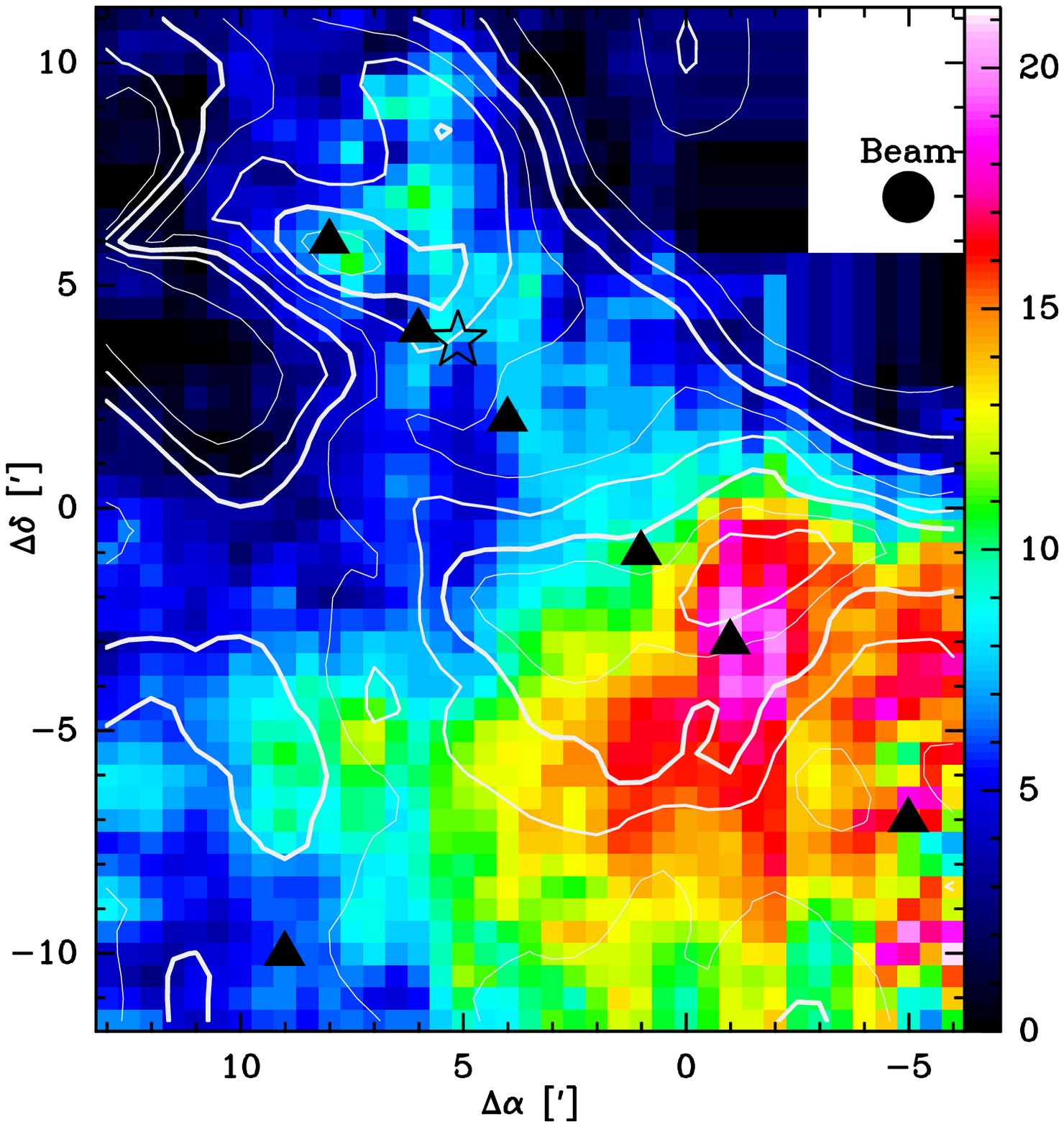}
\end{minipage}
\begin{minipage}[b]{6cm}
\includegraphics[width=6cm,angle=-90]{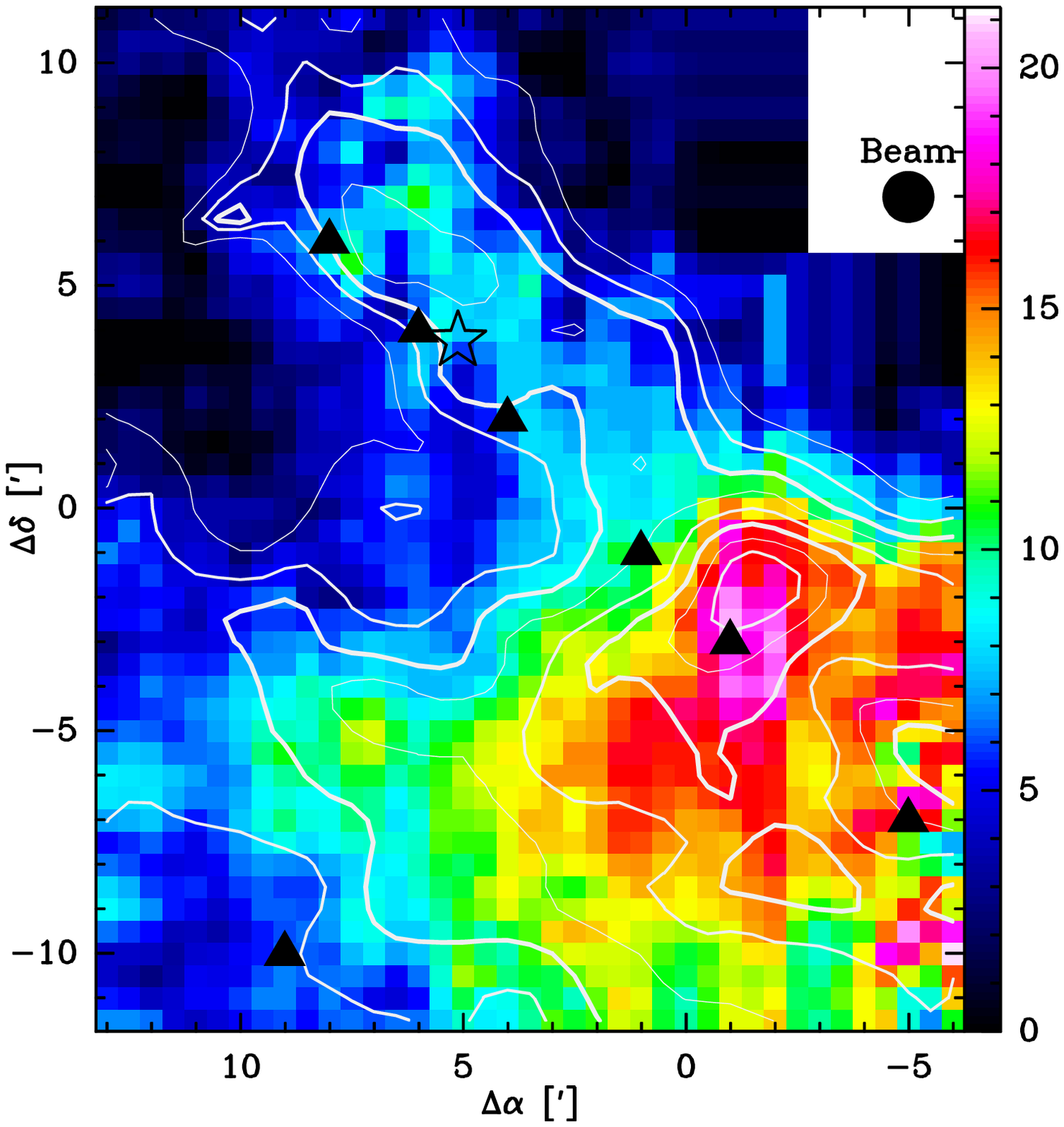}
\end{minipage}
}
 \caption{Velocity integrated intensities of [\ion{C}{i}] emission
(\emph{color}) overlayed with contours of (\emph{a}) $^{12}$CO
4--3, (\emph{b}) $^{12}$CO 1--0 and (\emph{c}) $^{13}$CO 1--0
integrated intensities at a common angular resolution of
70\arcsec. The center of the maps is at $\alpha$ =
03$^{h}$44$^{m}$10$^{s}$, $\delta$ = 32\degr06\arcmin (J2000). All
tracers are integrated from V$_{LSR}$ 2\,km\,s$^{-1}$ to
14\,km\,s$^{-1}$. Contours range between 20\% to 90\% with a step
of 10\% of the peak intensities that are 73\,K\,km\,s$^{-1}$ for
$^{12}$CO 4--3, 72\,K\,km\,s$^{-1}$ for $^{12}$CO 1--0 and
31\,K\,km\,s$^{-1}$ for $^{13}$CO 1--0. The seven filled triangles
indicates the positions where we carry out a detailed PDR analysis
later. The open pentacle symbol denotes the position of
HD\,281159.} \label{intenoverlay}
\end{figure*}

Figure~\ref{intenoverlay} presents maps of integrated intensities
of [\ion{C}{i}], $^{12}$CO 1--0, $^{13}$CO 1--0 and $^{12}$CO 4--3
observed in IC\,348. All maps are integrated over the velocity
range $V_{\rm LSR}$ = 2\,km\,s$^{-1}$ to 14\,km\,s$^{-1}$.

The energy of the upper level for the [\ion{C}{i}] 1--0 transition
is 24\,K and the critical density for this transition is
10$^{3}$\,cm$^{-3}$ for collisions with molecular H$_{2}$
\citep{sssfj91}. This implies that [\ion{C}{i}] 1--0 is easily
excited and the line is easily detectable even when emitted by
moderate density interstellar gas exposed to the average
interstellar radiation field. Previous observations have found the
[\ion{C}{i}] to be extended and well correlated with the low-$J$
CO emission \citep[see a review by][]{pzh96,sskksm03,mkrm06}. In
IC~348 the [\ion{C}{i}] emission peaks to the southwest of the
mapped region with a value of 24\,K\,km\,s$^{-1}$ and we detect
[\ion{C}{i}] emission over a large region with a homogeneous
intensity distribution at a level of 65\% of the  peak intensity.
The [\ion{C}{i}] emission is extended towards the east and
northeast of the mapped region. This is consistent with the
clumpy UV irradiated cloud scenario.

The $^{12}$CO 4--3 emission peaks at almost the same position as
[\ion{C}{i}] with an intensity of $\sim$ 66\,K\,km\,s$^{-1}$.
However, the morphology of the $^{12}$CO 4--3 emission is
different to that of [\ion{C}{i}]. We find secondary maxima in the
north and a steeper decay of the emission to the south. The
distribution of the $^{12}$CO 4--3 emission agrees very well with
that of $^{12}$CO 3--2 \citep{skobsm06}. We attribute the
difference in the intensity distributions of [\ion{C}{i}] and
$^{12}$CO 4--3 to the fact that $^{12}$CO 4--3 emission arises
from regions of higher temperature and density, while [\ion{C}{i}]
emission can arise also from embedded PDR surfaces with moderate
to low density within the molecular clouds.

The distribution of the $^{12}$CO 1--0 emission shows the same
global structure as the $^{12}$CO 4--3 emission, but it is
considerably more extended. This is consistent with the fact that
$^{12}$CO 1--0 traces lower temperature and density. The
morphologies of [\ion{C}{i}] and $^{13}$CO 1--0 appear to be very
similar, implying that they trace the same material and both lines
are effectively column density tracers.

\begin{figure*} %[h] \centering
\resizebox{\hsize}{!}{\includegraphics[width=\linewidth,angle=-90]{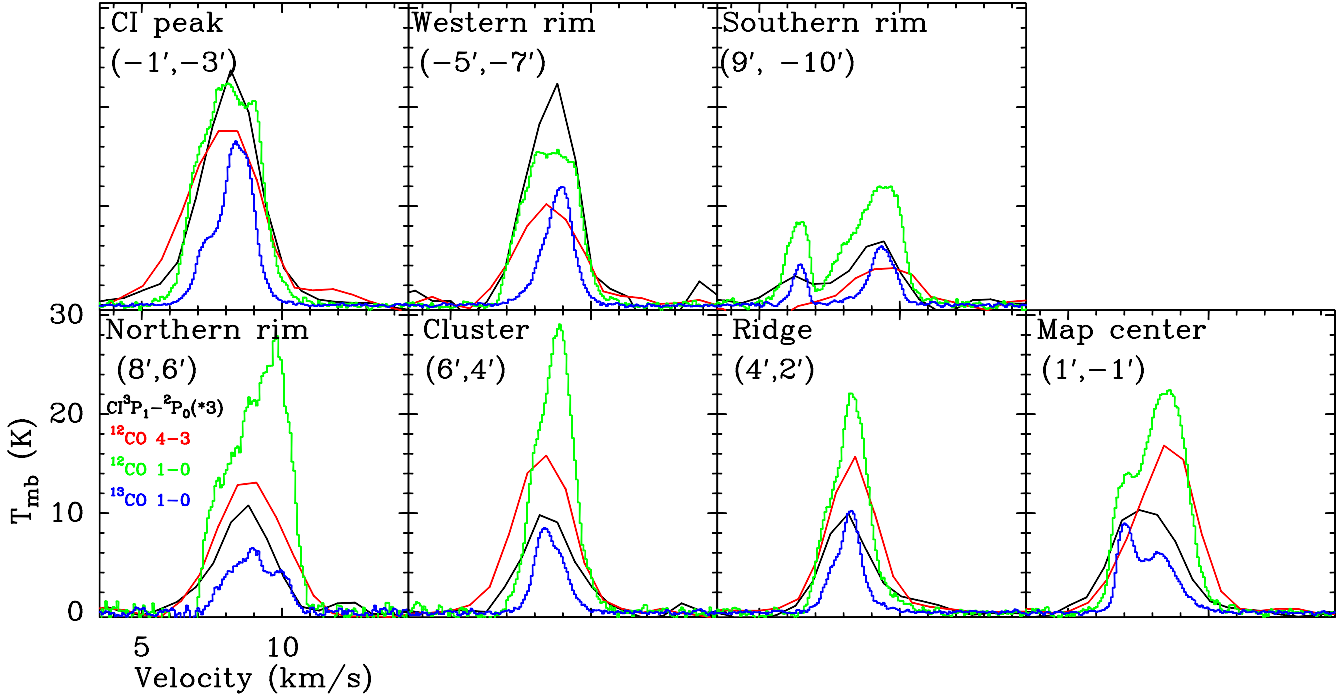}}
\caption{Spectra of the four different tracers on those six
positions along the cut from the northern edge of the cloud into
the cloud, and on the seventh position at the south of the cloud.
[\ion{C}{i}] spectra in all the positions have been multiplied by
3. All panels have the same x axis from 3.5\,km\,s$^{-1}$ to
14.5\,km\,s$^{-1}$ and the same y axis from -0.5\,K to 30.5\,K.}
\label{spectra}
\end{figure*}

We selected seven positions within IC\,348 for a more detailed
study. Six of the seven positions are oriented along a cut from
the northern edge of the cloud, past HD\,281159 and into the
clouds. The seventh position is at the southeast edge of the
cloud. Though emission at the seventh position (southern rim) is
weak, it is clearly detected. The selected positions vary in their
physical and chemical conditions. Spectra of [\ion{C}{i}],
$^{12}$CO 4--3, $^{12}$CO 1--0 and $^{13}$CO 1--0 at those seven
positions are displayed in Fig.~\ref{spectra}.

The [\ion{C}{i}] emission is centered at a velocity of $\sim$
8\,km\,s$^{-1}$. The lowest [\ion{C}{i}] main beam brightness
temperature ($\sim$ 2\,K) occurs at the southern rim
(9\arcmin,-10\arcmin), while the highest temperatures of 7.0\,K
and 6.7\,K are observed at the position of the [\ion{C}{i}] peak
(-1\arcmin,-3\arcmin) and the western rim (-5\arcmin,-7\arcmin),
respectively. The line widths of [\ion{C}{i}] spectra at the seven
positions range between 2.0\,km\,s$^{-1}$ and 2.6\,km\,s$^{-1}$.

The $^{12}$CO 4--3 main beam brightness temperature varies between
$\sim$ 3.6\,K at southern rim and 16.4\,K at [\ion{C}{i}] peak.
The line width of the $^{12}$CO 4--3 spectra ranges between
2.0\,km\,s$^{-1}$ and 2.7\,km\,s$^{-1}$.

Both the $^{12}$CO 1--0 and $^{13}$CO 1--0 spectra consist of two
velocity components: one is at $\sim$ 7\,km\,s$^{-1}$ and the
other lies at $\sim$ 8\,km\,s$^{-1}$. The two components are most
prominently visible at the southern rim, map center and the
[\ion{C}{i}] peak. For the 8\,km\,s$^{-1}$ component at the
[\ion{C}{i}] peak, the $^{13}$CO 1--0 spectrum peaks at the dip of
the $^{12}$CO 1--0 spectrum, which may indicate self-absorption
effect for this component.

\begin{figure} [h]
   \centering
   \includegraphics[width=7cm,angle=-90]{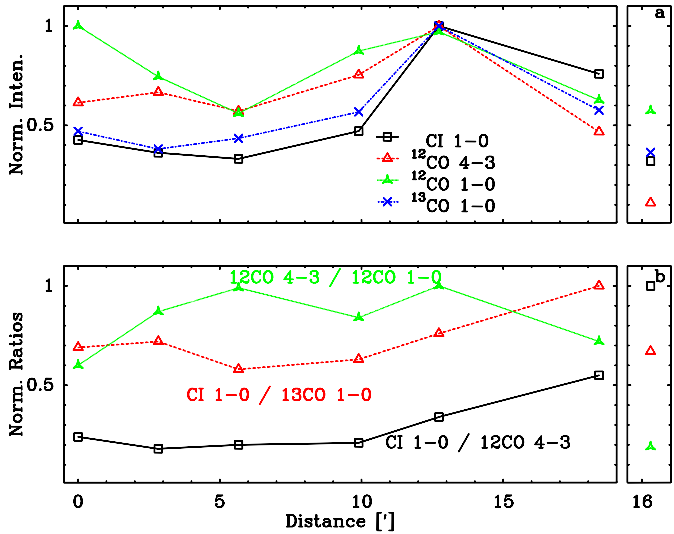}
      \caption{The top panel {\bf (a)} presents the normalized integrated intensities
      of [\ion{C}{i}], $^{12}$CO 4--3, $^{12}$CO 1--0 and $^{13}$CO 1--0 in the seven
      selected positions. The typical errors of the intensities are about 15\%,
      resulting in errors of the line ratios of about 21\%; {\bf (b)}
      displays integrated line intensity ratios. The x axes of all panels are the relative
      distance to the northern rim (8\arcmin,6\arcmin) position. The southern rim is plotted in a
      separate box.}
         \label{5poitions}
\end{figure}

Table~\ref{intensity} presents the integrated intensities and line
ratios at the  seven selected positions. Figure~\ref{5poitions}
shows the variation of the integrated intensities and their ratios
at these seven positions.

\begin{table*}
\caption{The observed integrated line intensities (in
erg\,s$^{-1}$\,sr$^{-1}$\,cm$^{-2}$) and line ratios at the seven
selected positions. The errors on the integrated intensity are
estimated to be $\sim$ 15\%.}
\label{intensity}      % is used to refer this table in the text
\centering                          % used for centering table
\begin{tabular}{lcrccccll}        % centered columns (4 columns)
\hline\hline                 % inserts double horizontal lines
Position            & ($\Delta\alpha$,$\Delta\delta$)& [\ion{C}{i}] & $^{12}$CO 4--3  & $^{12}$CO 1--0 & $^{13}$CO 1--0 & $\frac{\mathrm{[\ion{C}{i}] 1-0}}{\mathrm{^{12}CO 4-3}}$  & $\frac{\mathrm{[\ion{C}{i}] 1-0}}{\mathrm{^{13}CO 1-0}}$  & $\frac{\mathrm{^{12}CO 4-3}}{\mathrm{^{12}CO 1-0}}$ \\    % table heading
\cline{3-6}
                    &                                & \multicolumn{4}{c}{[10$^{-8}$ erg\,s$^{-1}$\,sr$^{-1}$\,cm$^{-2}$]}                                                          &                                         &                                         &                                   \\
\hline
Northern rim        & ( 8\arcmin, 6\arcmin)          &  96.56            & 336.43          & 9.59            & 1.74                                                                 & 0.29                                    & 55.51                                   & 35.07                             \\
Cluster             & ( 6\arcmin, 4\arcmin)          &  81.99            & 365.30          & 7.14            & 1.41                                                                 & 0.22                                    & 58.01                                   & 51.15                             \\
Ridge               & ( 4\arcmin, 2\arcmin)          &  75.09            & 313.20          & 5.37            & 1.61                                                                 & 0.24                                    & 46.59                                   & 58.38                             \\
map Center          & ( 1\arcmin,-1\arcmin)          & 106.79            & 413.27          & 8.38            & 2.10                                                                 & 0.26                                    & 50.85                                   & 49.33                             \\
$[\ion{C}{i}]$ peak & (-1\arcmin,-3\arcmin)          & 226.61            & 548.04          & 9.31            & 3.70                                                                 & 0.41                                    & 61.23                                   & 58.85                             \\
Western rim         & (-5\arcmin,-7\arcmin)          & 172.05            & 256.15          & 6.01            & 2.13                                                                 & 0.67                                    & 80.61                                   & 42.61                             \\
Southern rim        & ( 9\arcmin,-10\arcmin)         &  72.71            &  60.21          & 5.50            & 1.35                                                                 & 1.21                                    & 54.02                                   & 10.95                             \\

\hline                                   %inserts single line
\end{tabular}
\end{table*}

In Fig.~\ref{5poitions}a we see that the intensities of
[\ion{C}{i}], $^{13}$CO 1--0  and $^{12}$CO 4--3 lines show
similar trends, only the falling off of the intensity of $^{12}$CO
4--3 is somewhat less drastic than [\ion{C}{i}], $^{13}$CO 1--0.
As opposed to the single peak seen in [\ion{C}{i}], $^{13}$CO 1--0
and $^{12}$CO 4--3, the  $^{12}$CO 1--0 intensity profile shows
two peaks with comparable intensities, one at the northern rim and
the other at the position of the [\ion{C}{i}] peak.

Three independent line ratios, i.e., [\ion{C}{i}] / $^{12}$CO
4--3, [\ion{C}{i}] / $^{13}$CO 1--0 and $^{12}$CO 4--3 / $^{12}$CO
1--0, are presented in Table~\ref{intensity} and
Fig.~\ref{5poitions}b (in erg\,s$^{-1}$\,sr$^{-1}$\,cm$^{-2}$).
The largest [\ion{C}{i}] / $^{12}$CO 4--3 ratio (1.21) occurs at
the southern rim and the second (0.67) and third (0.41) largest
are at the western rim and [\ion{C}{i}] peak, while the ratios at
the other four positions are $\sim$ 0.25. When comparing with
other star forming regions like W\,3\,Main, S\,106 and Orion A
etc. \citep[see Table~3 by][]{kjmsbs04}, the [\ion{C}{i}] /
$^{12}$CO 4--3 ratio at the other five positions are within the
range found in those star forming regions except for the western
and southern rim. The highest ratio at the southern rim is close
to the ratio found at the center of M\,51 and the nucleus position
of NGC\,4826 \citep{ib02}. The $^{12}$CO 4--3 / $^{12}$CO 1--0
ratio peaks at [\ion{C}{i}] peak and decreases for the positions
further away with increasing distance. The [\ion{C}{i}] /
$^{13}$CO 1--0 ratios are rather constant except for the western
rim.

%________________________________________________________________

\section{LTE Analysis}\label{sec_lte}

\begin{figure} [h]
\centering

\includegraphics[width=8cm,angle=-90]{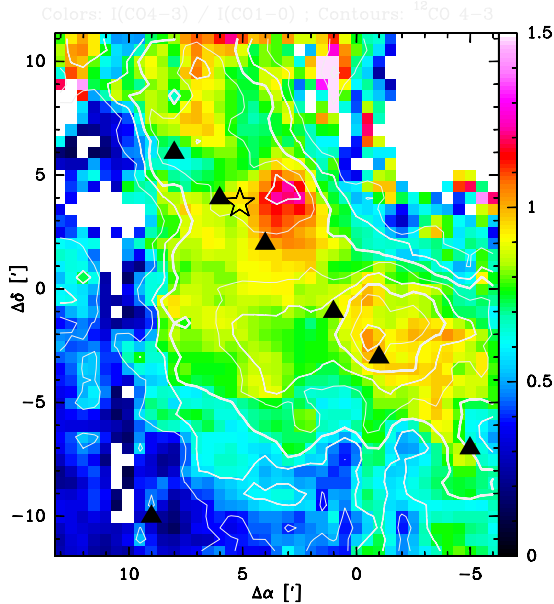}

\caption{The color plot represents the $^{12}$CO 4--3 / $^{12}$CO
1--0 line intensity ratio (in units of K\,km\,s$^{-1}$) observed
in the IC\,348 cloud. The contours show the $^{12}$CO 4--3 line
intensity spaced at 10\% intervals from 20\% to 90\% relative to
the peak intensity of 73\,K\,km\,s$^{-1}$. White pixels in the
figure denote positions where either of the line intensities falls
below the 3 $\sigma$ threshold.} \label{lte}
\end{figure}

Figure~\ref{lte} shows a map of the $^{12}$CO 4--3 / $^{12}$CO 1--0
line ratio in terms of line integrated temperatures (K km s$^{-1}$)
overlaid by $^{12}$CO 4--3 integrated intensities. The ratio has
its minimum of $\sim$ 0.3 at the edges of the cloud; ratios of
$\sim$ 0.9 are found at the $^{12}$CO 4--3 peaks; maximum ratios
between 1.1 and 1.5 occur close to HD\,281159.

For a rough estimate of the column densities and for an easy
comparison to other \CI{} studies \citep[e.g.][]{little1994,
plume2000,bensch2003} we first apply a Local Thermodynamic
Equilibrium analysis. Assuming LTE and optically thick emission,
the $^{12}$CO 4--3 / $^{12}$CO 1--0 ratio of 0.3 corresponds to an
excitation temperature $T_\mathrm{ex}$ of $\sim$ 10\,K. The ratio
grows to 0.85 for $T_\mathrm{ex} \sim 50$\,K and remains almost
constant at even higher temperatures. Considering the calibration
uncertainty, we set $T_\mathrm{ex}$ to 50\,K in all the positions
where the ratio falls above 0.85. We note that [\ion{C}{i}] and
$^{12}$CO often have different excitation temperatures. However,
$N$(C)$_\mathrm{LTE}$ only changes by 20\% when the assumed
excitation temperature varies between 20 and 150\,K. For the first
order estimation, it is therefore reasonable to use the $^{12}$CO
excitation temperature also for [\ion{C}{i}]. With the $^{12}$CO
excitation temperatures and the assumption of optically thin
[\ion{C}{i}] and $^{13}$CO 1--0 emission, we then obtain the
column density of C, $N$(C)$_\mathrm{LTE}$, and $^{13}$CO,
$N$($^{13}$CO)$_\mathrm{LTE}$, from the integrated line
intensities. Using the relative abundance ratio
[$^{12}$CO]/[$^{13}$CO] of 65 \citep{lp90} we translate the
$^{13}$CO column densities into CO column densities. H$_{2}$
column densities are derived from the [CO]/[H$_{2}$] abundance
ratio of 8\,10$^{-5}$ \citep{flw82}. The results are summarized in
Table~\ref{lteta}.

\begin{table*}
\caption{Results of the LTE analysis. The excitation temperature
$T_\mathrm{ex}$ is listed in Column (2); Column (3) - (5) present
C, CO and H$_{2}$ column densities; the C/CO ratio is listed
in Column (6). The uncertainties of the column densities are derived
by varying the integrated intensities by $\pm$15\%.}
\label{lteta}      % is used to refer this table in the text
\centering                          % used for centering table
\begin{tabular}{lrrrrc}        % centered columns (4 columns)
\hline\hline                 % inserts double horizontal lines
number               & $T_\mathrm{ex}$       & $N$(C)$_\mathrm{LTE}$  & $N$(CO)$_\mathrm{LTE}$ & $N$(H$_{2}$)$_\mathrm{LTE}$ & C/CO$_\mathrm{LTE}$ \\
                     &    K           & 10$^{16}$[cm$^{-2}$]   & 10$^{17}$[cm$^{-2}$]   & 10$^{21}$[cm$^{-2}$]        &                     \\
\hline        %inser
Northern rim         & 18$^{47}_{15}$ & 12.08$^{19.71}_{9.27}$ &  9.56$^{22.84}_{7.16}$ & 12.50$^{28.55}_{8.95}$      & 0.13$^{0.28}_{0.04}$     \\
Cluster              & 50$_{20}$      &  9.61$^{11.17}_{7.77}$ & 16.99$^{19.54}_{7.33}$ & 21.24$^{24.43}_{9.16}$      & 0.06$^{0.15}_{0.04}$     \\
Ridge                & 50$_{28}$      &  9.04$^{10.46}_{7.32}$ & 19.37$^{23.03}_{10.43}$& 24.21$^{28.78}_{13.04}$     & 0.05$^{0.10}_{0.03}$     \\
Map center           & 48$^{50}_{18}$ & 13.09$^{16.00}_{10.68}$& 24.39$^{29.03}_{10.25}$& 30.49$^{36.29}_{12.82}$     & 0.05$^{0.16}_{0.04}$     \\
$[\ion{C}{i}]$ peak  & 50$_{29}$      & 26.85$^{30.88}_{21.71}$& 44.47$^{51.14}_{24.56}$& 55.59$^{63.93}_{30.70}$     & 0.06$^{0.13}_{0.04}$     \\
Western rim          & 28$^{50}_{15}$ & 19.72$^{28.30}_{17.61}$& 16.25$^{29.99}_{9.52}$ & 20.31$^{37.49}_{11.90}$     & 0.12$^{0.30}_{0.06}$     \\
Southern rim         &  7$^{8}_{6}$   & 37.50$^{72.57}_{21.96}$&  6.66$^{8.54}_{5.40}$  &  8.32$^{10.67}_{6.76}$      & 0.56$^{1.34}_{0.26}$     \\
\hline                                   %inserts single line
\end{tabular}
\end{table*}

\begin{figure}[h]
\centering
\includegraphics[width=8cm,angle=-90]{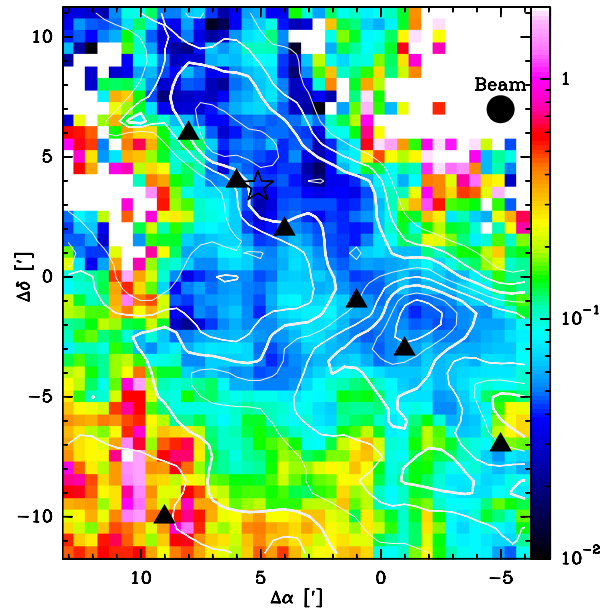}
\caption{The C/CO column density ratio overlaid by $^{13}$CO 1--0
integrated intensities contours. The contour levels are spaced at
10\% intervals from 20\% to 90\% relative to the peak intensity of
31\,K\,km\,s$^{-1}$. White pixels in the figure denote positions
where the $^{13}$CO 1--0 intensities fall below the 3 $\sigma$
threshold.} \label{nco}
\end{figure}

The resulting map of C/CO column density ratios is presented in
Fig.~\ref{nco}. In most of the cloud, the C/CO ratio falls below
0.1. Higher values (up to 1.5) occur at the rim of the cloud where
the $^{13}$CO 1--0 emission is more diffuse. Values as low as 0.02
are found at the southwest of the cloud and along the ridge to the
northeast. As $^{13}$CO 1--0 roughly traces the H$_{2}$ column
density, this plot indicates an anti-correlation between the C/CO
ratio and the H$_{2}$ column density. This is consistent with the
conclusion of \citet{mkrm06} that the [\ion{C}{i}] line is not a
tracer of optical extinction, total H$_{2}$ column density or
total mass in the beam.

The seven positions, for which we perform a detailed analysis
within the context of a PDR model, exhibit C/CO column density
ratios between about 0.05 and 0.5, i.e. they cover a relatively
wide range (see Table~\ref{lteta}). The ratios measured within the
full map cover almost the full range cited so far in the
literature with values as low as 0.03 in parts of G34.3
\citep{little1994} and as high as 1.5 seen in the Polaris Flare
\citep{bensch2003}.

%________________________________________________________________

\section{PDR Modelling}\label{sec_pdr}

In the next step we attempt to obtain a self-consistent model for
the physical and chemical structure of the cloud matching the
observed C/CO ratios. We will use a clumpy PDR model to
self-consistently compute the abundance of the species and their
excitation determining the line strength. The model is constrained
by the strength of the UV field, which can be independently
estimated, and by the known geometry, provided by the complex,
filamentary, and clumpy structure of the region.

\subsection{FUV intensity}\label{sec_fuv}

\subsubsection{Estimate from the FIR continuum}

\begin{figure} [h]
   \centering
   \includegraphics[width=8cm,angle=-90]{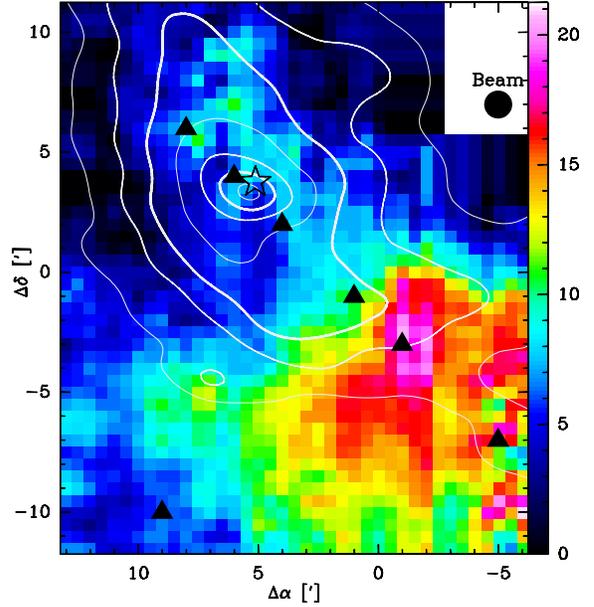}
      \caption{FUV intensity distribution (contours) in IC\,348
      estimated from the IRAS FIR fluxes, overlayed on the
      [\ion{C}{i}] integrated intensities (colors).
      The FUV intensity $\chi_\mathrm{FIR}$ contour levels run from 5, 10,
      20 to 100 by steps of 20 Draine units, $\chi_{0}$.
              }
         \label{fuv}
\end{figure}

We use HIRES processed 60 and 100~$\mu$m IRAS data \citep{afm90}
to generate a far-infrared intensity map $I_\mathrm {FIR}$ in
IC\,348. It represents the far-infrared intensity between
42.5\,$\mu$m and 122.5\,$\mu$m as measured by the filter curves of
the two IRAS data sets \citep{nyd98}. Following
\citet[][]{kwhl99}, we assume that the total FUV flux,
$\chi_\mathrm{FIR}$, absorbed by the dust grains is re-radiated in
the far-infrared. In this way we estimate the FUV field from the
far-infrared field \citep[cf.][]{kmbggisw05}: $\chi_\mathrm{FIR} /
\chi_{0} = 4\pi~I_\mathrm{FIR}$, where $\chi_{0} = 2.7 \times
10^{-3}$\,erg\,s$^{-1}$\,cm$^{-2}$ \citep{dra78,db96}. The derived
spatial distribution of the FUV intensity is presented in
Fig.~\ref{fuv}. In the mapped region, the FUV field
$\chi_\mathrm{FIR}$ varies between about 1 and 100 Draine units.
For the seven selected positions $\chi_\mathrm{FIR}$ is listed in
Table~\ref{pdrtable}.

\subsubsection{Estimate from the stellar radiation}

The primary source of UV radiation in the cloud is HD\,281159.
Assuming that the star is a black body at an effective temperature
$T_\mathrm{eff}$ corresponding to its spectral type, we can
estimate the FUV flux independently \citep{ksw96,bkds00}. A
B\,5\,V star has an effective temperature $\log(T_\mathrm{eff}) =
4.182$ and a total luminosity $L$ of $\log(L/L_{\odot}) = 2.681$
\citep{dn87}. The FUV luminosity $L_\mathrm{FUV}$ and the FUV flux
$\chi_\mathrm{FUV}$ are defined as
\begin{equation}
 L_\mathrm{FUV} = L \cdot
 \frac{\Phi_\mathrm{UV}}{\Phi_\mathrm{total}}, ~~{\rm{and}}~~
 \chi_\mathrm{FUV} = \frac{L_\mathrm{FUV}}{4\pi d^{2}},
\end{equation}
where $\Phi_\mathrm{UV}$ is the FUV flux between 6 eV and 13.6 eV,
$\Phi_\mathrm{total}$ is the total flux, and $d$ is the distance
to the UV source. We compute the distance $d$ in two ways: for the
point close to HD\,281159 we use the optical radius of the cluster
IC\,348 of about 0.37\,pc \citep{her98} as the minimum distance
between star and cloud. For all other points we assume that star
and cloud are located in the same plane so that the distance is
directly given by the observed separation. Finally we normalize
the FUV field to units of $\chi_{0}$. The resulting FUV fields,
$\chi_\mathrm{star}$, are also listed in Table~\ref{pdrtable} for
the seven selected positions.

We see that both methods provide consistent FUV field values.
Because of the uncertainty of the distance between star and cloud
we consider the values $\chi_\mathrm{FIR}$, derived from the HIRES
data, somewhat more reliable so that we will use them in the
following. The FUV field $\chi$ at the seven positions ranges from
about 1 to 90 Draine units.

%________________________________________________________________

\subsection{Clumpy PDR scenarios}

Molecular clouds are known for their highly filamentary and
clumpy, often self-similar structure over a wide range of scales
\citep{dht01,hbs98,ssghjb88}. The observed structure further
breaks up into substructures with every step of increased spatial
resolution \citep{fp96,bso01}. For IC\,348, the self-similar
structure was directly measured in terms of the $\Delta$-variance
by \citet{skobsm06}. In such a cloud, the UV field can penetrate
deeply into the cloud forming PDRs at many internal surfaces. The
cloud is basically filled by surfaces \citep{orcs07}. This
explains the observed [\ion{C}{i}] and even [\ion{C}{ii}] emission
deep within molecular clouds \citep{ssghjb88,hjgs91,sjgg93,mt93}.
We can simulate such a structure by an ensemble of individual
clumps, each of them being small compared to the overall cloud,
leaving enough ``empty" regions to permit a deep penetration of
the UV field \citep{sbhoz98}.

In the following we test two different clump ensembles with
respect to their ability to reproduce the observed line
intensities.

\subsubsection{Modelling of individual clumps}

All individual clumps in the PDRs are modelled using the spherical
KOSMA - $\tau$ PDR model \citep{sss96,rojs06, roellig07}. The
model computes the chemical and temperature structure of a clump
illuminated by an isotropic FUV radiation field and cosmic rays.
The emission from the model is calculated as a function of the gas
density, FUV radiation field and mass of the clumps (implicitly
specifying the clump size). The model clumps are assumed to have a
power-law density profile of $n$(r) $\propto$ $n_{0}$$r^{-1.5}$
for 0.2 $\leq$ $r/r_\mathrm{cl}$ $\leq$ 1 and $n$(r) = const. for
$r/r_\mathrm{cl}$ $\leq$ 0.2. $n_{0}$ is the density at the clump
surface called \emph{clump density}, which is about half of the
clump averaged density $\langle n \rangle$: $n_{0}$ $\sim$
$\frac{1}{1.91}$ $\langle n \rangle$.

The PDR models are pre-computed on a regular grid with equidistant
logarithmic steps. The FUV field $\chi$ covers the range from
$10^{0}, 10^{0.5}, ..., 10^{5.5}, 10^{6.0}$ Draine units; The
clump densities range from $10^{2.0}, 10^{2.5}, ..., 10^{5.5},
10^{6.0}$ cm$^{-3}$; and the clump masses $M_\mathrm{cl}$ cover
$10^{-3.0}, 10^{-2.5}, ..., 10^{1.5}, 10^{2.0}$ M$_{\odot}$.

\subsubsection{Ensemble of identical clumps (Ensemble 1)}

\begin{figure*}%[h]
\centering
\mbox{
\begin{minipage}[b]{8cm}
\includegraphics[width=7cm,angle=-90]{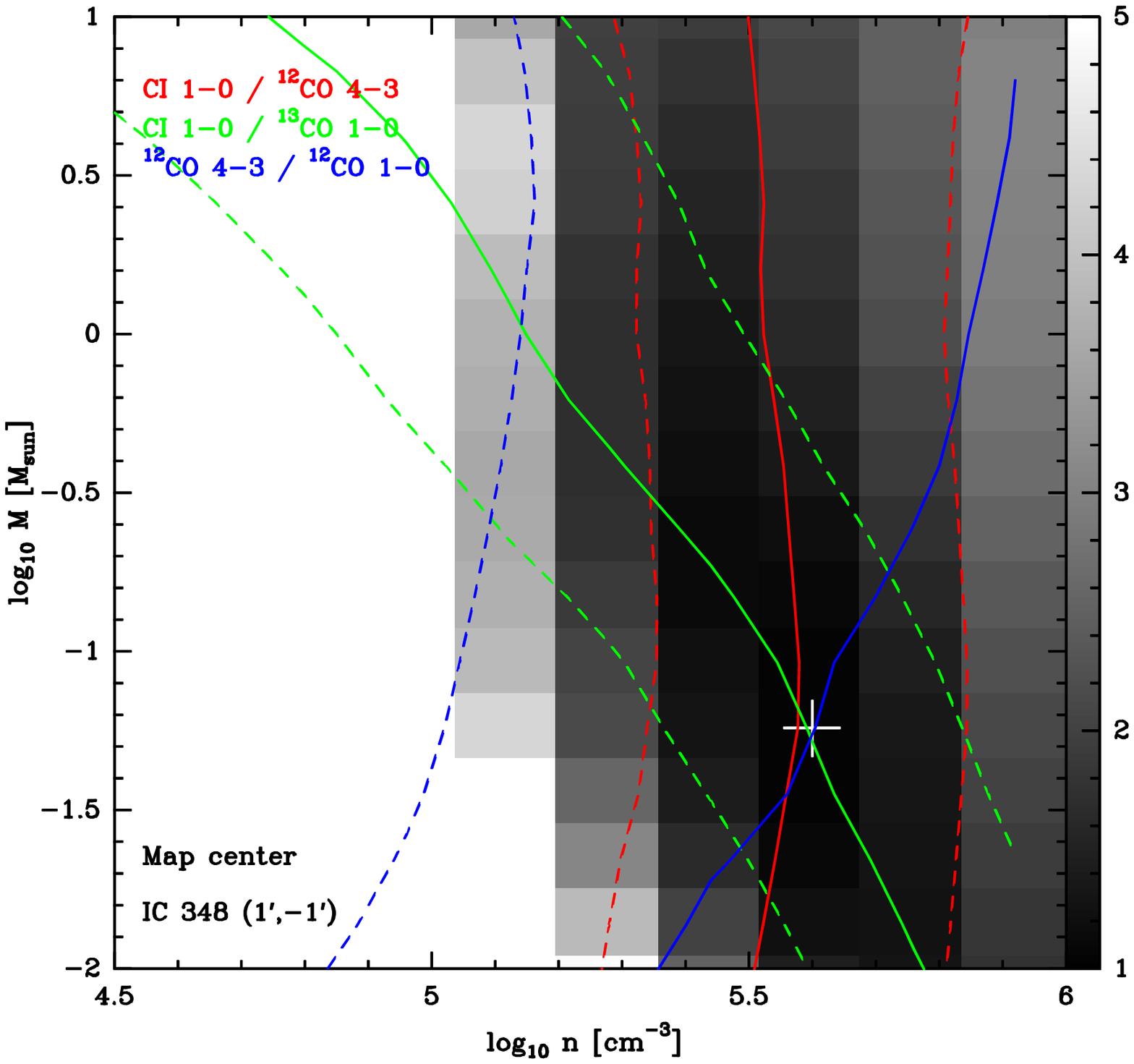}
%\centerline{a}
\end{minipage}
\begin{minipage}[b]{8cm}
\includegraphics[width=7cm,angle=-90]{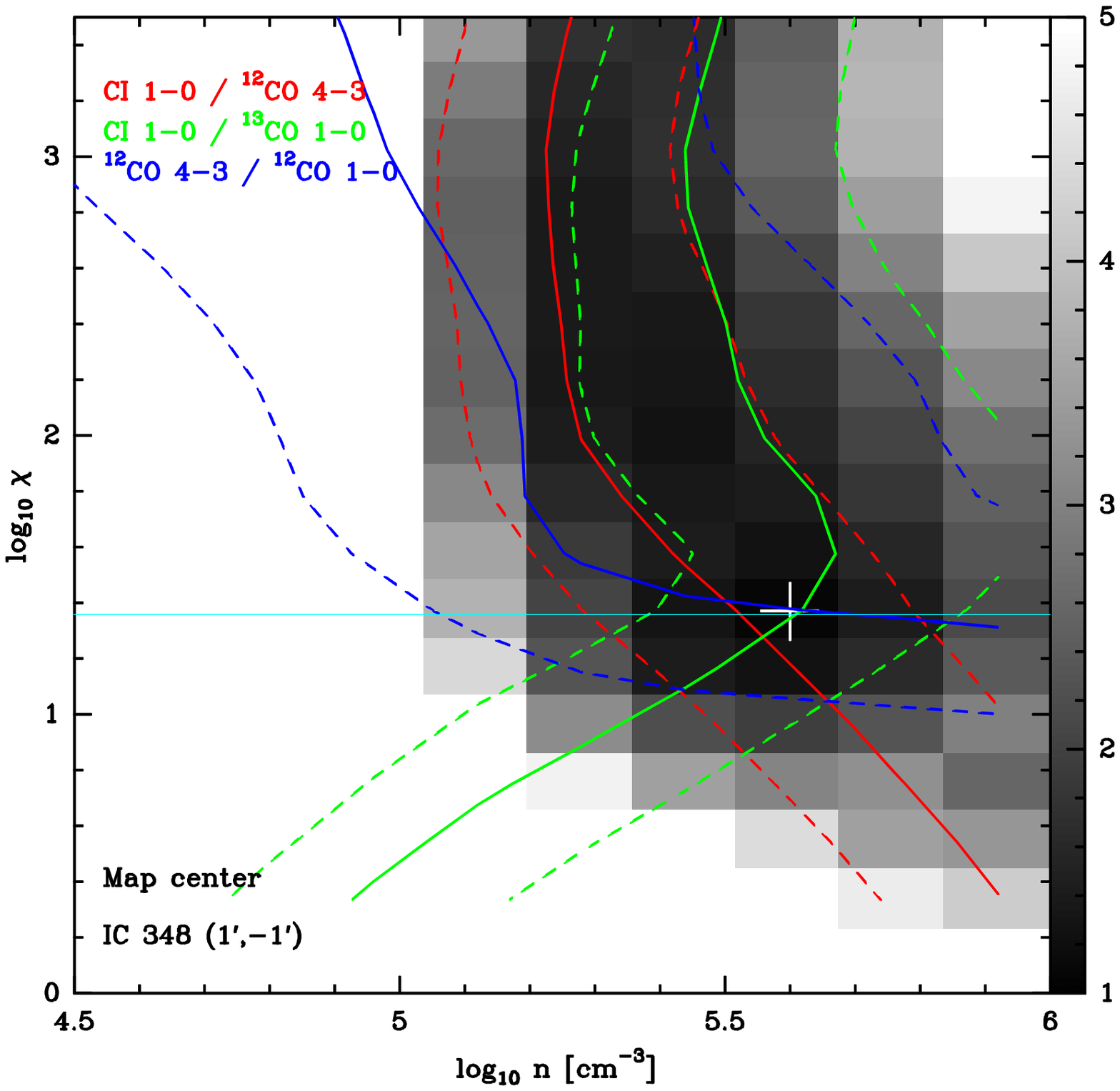}
%\centerline{b}
\end{minipage}
} \caption{Comparison of the observed line intensity ratios
[\ion{C}{i}]/$^{12}$CO 4--3, [\ion{C}{i}]/$^{13}$CO 1--0 and
$^{12}$CO 4--3/$^{12}$CO 1--0 at (1\arcmin, $-1$\arcmin) with the
single clump KOSMA - $\tau$ PDR model calculations. The left
figure shows the fit of clump mass and density at a fixed FUV
field $\chi$ of 10$^{1.5}$; the right one presents the fit of FUV
field and clump density at a fixed clump mass of 10$^{-1}$
M$_{\odot}$. The central solid contours represent the observed
intensity ratios, the outer dashed two give those for a 20\%
uncertainty. The grey-scale images indicate the reduced $\chi^{2}$
of the fit. The position of the minimum reduced $\chi^{2}$ is
marked by a white cross. The horizontal light blue line in the
right panel shows the FUV flux $\chi_\mathrm{FIR}$ from the HIRES
data.} \label{pdra}
\end{figure*}

In a first approach we will assume that the cloud is composed of a
collection of identical clumps, so that the full ensemble is
characterized by the parameter set of the individual clumps and
their number only. To determine the properties of the individual
clumps we perform a $\chi^{2}$ fit of the three line intensity
ratios given in Table~\ref{intensity}, using a spherical PDR model
in a three-dimensional parameter space spanned by the FUV field,
the clump mass, and density. The results for the seven selected
positions are summarized in Table~\ref{pdrtable}.
Figure~\ref{pdra} presents an example of the $\chi^{2}$ fitting
for the map center (1\arcmin, $-1$\arcmin). The left panel shows a
cut through the parameter space for a fixed UV field,
demonstrating the fit of clump mass M$_{cl}$ and clump density
$n_{0}$. The right panel presents the corresponding fit for the
FUV field $\chi$ and the clump density $n_{0}$. We see that all
line ratios are good density tracers, but hardly constrain the
clump mass as expected. The clump density has to fall between 2.5
and 5.0\,10$^{5}$ cm$^{-3}$. But the reduced $\chi^{2}$ only
changes between 1 and 2.5 when the clump mass varies from
10$^{-2.0}$ to 10$^{1.0}$ M$_{\odot}$. This general behaviour
applies to all positions analyzed. The clump densities at the
seven positions vary between 4.4\,10$^{4}$ to
4.3\,10$^{5}$\,cm$^{-3}$. The highest clump density occurs at
positions of the ridge and the map center. The clump masses are
not well constrained, the best fitting values fall between 0.1 and
0.4\,M$_{\odot}$. The fitted FUV flux $\chi_{\rm PDR}$ agrees with
$\chi_{\rm FIR}$ derived from the FIR continuum in most positions.
The maximum deviations by a factor three correspond to the maximum
deviation that we got comparing the FIR continuum with the stellar
radiation estimate.

\begin{table*}
\caption{Physical parameters at selected positions in
Fig~\ref{intenoverlay}, obtained by fitting the observed intensity
ratios to the {\bf single clump} KOSMA - $\tau$ PDR models.
Columns (2) - (5) list the fitted clump densities, clump mass,
clump radius and FUV field; Column (6) is the minimum reduced
$\chi^{2}$ of the PDR fit; the FUV field derived from the FIR
continuum $\chi_\mathrm{FIR}$ and that from the stellar radiation
$\chi_\mathrm{star}$ are in Column (7) and (8); The distance to
HD\,281159 used for the calculation of $\chi_\mathrm{star}$is
listed in Column (9).}
\label{pdrtable}      % is used to refer this table in the text
\centering                          % used for centering table
\begin{tabular}{cccccrrcc}        % centered columns (4 columns)
\hline\hline                 % inserts double horizontal lines
Position                                 & $n$                   & $M_\mathrm{cl}$ & R$_\mathrm{cl}$   & $\chi_\mathrm{PDR}$& $\chi^{2}_{min}$ & $\chi_\mathrm{FIR}$ & $\chi_\mathrm{star}$ & $d$   \\    % table heading
                                         & [10$^{5}$cm$^{-3}$]   & M$_{\odot}$     & pc                &                    &                  &                     &                      & pc    \\
\hline                        % inserts s
Northern rim                             & 1.91                  & 0.2             & 0.01              & 15                 & 1.38             & 39                  &  40                  & 0.37  \\
Cluster                                  & 2.75                  & 0.1             & 0.01              & 98                 & 1.05             & 84                  &  40                  & 0.37  \\
Ridge                                    & 4.25                  & 0.4             & 0.01              & 38                 & 1.23             & 46                  &  40                  & 0.37  \\
Map center                               & 3.98                  & 0.1             & 0.01              & 23                 & 1.04             & 23                  &  13                  & 0.64  \\
$[\ion{C}{i}]$ peak                      & 1.91                  & 0.1             & 0.02              & 23                 & 1.96             & 11                  &   7                  & 0.90  \\
Western rim                              & 0.63                  & 0.2             & 0.02              & 20                 & 2.00             &  6                  &   3                  & 1.42  \\
Southern rim                             & 0.44                  & 0.4             & 0.02              &  2                 & 2.31             &  1                  &   3                  & 1.29  \\
\hline                                   %inserts single line
\end{tabular}
\end{table*}

The total number of clumps in the beam $N_\mathrm{ens}$ is
calculated by comparing the observed absolute line integrated
intensities with the models, correcting for the beam filling
factor.\footnote{By using line integrated intensities, a smaller
internal velocity distribution in the clumps relative to the total
observed line width will be automatically corrected by a higher
clump number.} A 70\arcsec\ beam corresponds to a diameter of
0.10\,pc at the distance of IC\,348. The clump size obtained from
the best fit models found in the previous section ranges from 0.01
to 0.02\,pc. The beam filling factor $\eta_\mathrm{bf,i}$ of each
clump is calculated as $\eta_\mathrm{bf,i} = \Omega_\mathrm{cl} /
\Omega_\mathrm{beam}$, where $\Omega_\mathrm{cl}$ and
$\Omega_\mathrm{beam}$ are the solid angles of a clump and the
beam.  $\Omega_\mathrm{cl} = \pi R_\mathrm{cl}^{2} / D^{2}$ is the
solid angle of the individual clump. $D$ is the distance of the
cloud; assuming a Gaussian beam shape, $\Omega_\mathrm{beam}$ is
computed as $\Omega_\mathrm{beam} = \pi \theta_\mathrm{beam}^{2} /
4\ln2$, where $\theta_\mathrm{beam}$ is the FWHM of our beam. The
intensity correction for a single clump is
$I^\mathrm{\prime}_\mathrm{i} = I_\mathrm{mb,i} /
\eta_\mathrm{bf,i}$, where $I_\mathrm{mb,i}$ is the beam averaged
line intensity for that single clump. Since [\ion{C}{i}] is the
most optically thin one among our four tracers, we use its line
intensity to derive the number of clumps in the beam
$N_\mathrm{ens}$ (see Table~\ref{models}) by dividing the observed
intensities $I_\mathrm{[\ion{C}{i}]obs}$ by the intensities for a
single clump $I_\mathrm{[\ion{C}{i}]mod,i}$

\begin{equation}
N_\mathrm{ens} =
\frac{I_\mathrm{[\ion{C}{i}]obs}}{I_\mathrm{[\ion{C}{i}]mod,i}
\cdot \eta_\mathrm{bf,i}} =
\frac{I_\mathrm{[\ion{C}{i}]obs}}{I_\mathrm{[\ion{C}{i}]mod,i}}
\times \frac{\Omega_\mathrm{beam}}{\Omega_\mathrm{cl}}.
\end{equation}

\begin{table*}
\caption{Results of ensemble of identical clumps, {\bf Ensemble
case 1}. The mass and FUV field for each single clump are listed
in Column (2) and (3); Column (4) - (6) present the line intensity
ratios between the ensemble models and observations of $^{12}$CO
4--3, $^{12}$CO 1--0 and $^{13}$CO 1--0; The C, CO and H$_{2}$
column densities from the ensembles are presented in Column (7),
(8) and (9); Column (10) is the C/CO ratios. The total number of
the clumps in the ensemble and ensemble mass are listed in Column
(11) and (12); Column (13) lists the beam filling factor of each
clump. The total ensemble mass $M_\mathrm{ens}$ listed in the
table is computed as $M_\mathrm{ens} = N_\mathrm{ens} \times
M_\mathrm{cl}$, where $M_\mathrm{cl}$ is the mass of a clump.}
\label{models}      % is used to refer this table in the text
\centering                          % used for centering table
\begin{tabular}{lcrcccrrrcrcc}        % centered columns (4 columns)
\hline\hline                 % inserts double horizontal lines
Position            & $M_\mathrm{cl}$  & $\chi_\mathrm{en1}$  & $R_\mathrm{^{12}CO 4-3}$ & $R_\mathrm{^{12}CO 1-0}$ & $R_\mathrm{^{13}CO 1-0}$ & $N$(C)$_\mathrm{ens}$ & $N$(CO)$_\mathrm{ens}$ & $N$(H$_{2}$)$_\mathrm{ens}$ & C/CO$_\mathrm{ens}$ & $N_\mathrm{ens}$ & $M_\mathrm{ens}$ & $\eta_\mathrm{bf,i}$ \\
                    & M$_{\odot}$      &                     &                          &                          &                          & 10$^{16}$[cm$^{-2}$]  & 10$^{17}$[cm$^{-2}$]   & 10$^{21}$[cm$^{-2}$]        &                     &                  & M$_{\odot}$      &                    \\
\hline
Northern rim        & 0.32             &  10                 & 0.63                     & 0.67                     & 0.87                     & 13.97                 & 18.58                  &  7.72                       & 0.075               &  4.59            &  1.47             &  0.17              \\
Cluster             & 0.10             & 100                 & 1.07                     & 0.99                     & 1.13                     & 10.02                 & 21.56                  &  8.90                       & 0.046               & 16.75            &  1.68             &  0.04              \\
Ridge               & 0.32             &  32                 & 0.95                     & 1.13                     & 0.95                     &  8.90                 & 29.49                  & 11.29                       & 0.030               &  6.72            &  2.15             &  0.08              \\
Map center          & 0.10             &  32                 & 0.94                     & 1.05                     & 0.97                     & 13.25                 & 34.24                  & 13.05                       & 0.039               & 24.56            &  2.46             &  0.04              \\
$[\ion{C}{i}]$ peak & 0.10             &  32                 & 0.94                     & 1.59                     & 0.83                     & 31.54                 & 29.88                  & 13.14                       & 0.106               & 24.74            &  2.47             &  0.08              \\
Western rim         & 0.10             &  10                 & 0.78                     & 1.43                     & 0.71                     & 24.52                 &  9.33                  &  6.10                       & 0.263               & 11.51            &  1.15             &  0.17              \\
Southern rim        & 0.32             &   1                 & 0.84                     & 0.81                     & 0.74                     & 14.62                 &  6.28                  &  3.03                       & 0.233               &  5.70            &  1.82             &  0.17              \\
\hline                                   %inserts single line
\end{tabular}
\end{table*}

Then we use $N_\mathrm{ens}$ to calculate the ratios $R$ of line
intensities between model ensemble and observations for the other
tracers:

\begin{equation}
R = \frac{N_\mathrm{ens} \cdot
I_\mathrm{mod,i}\cdot \eta_\mathrm{bf,i}}{I_\mathrm{obs}}.
\end{equation}

The $R$ values are listed in Table~\ref{models}.
This allows easy judgment of the quality of the model fit. Considering
the calibration uncertainty, a perfectly fitting model yields a ratio
$R$ of 1$\pm$0.15. All
three remaining tracers from Ensemble 1 have a good agreement
between the modelled and observed absolute line intensities.

\subsubsection{Clumps with a mass and size spectrum (Ensemble 2)}

Representing the clumpy structure of the cloud by an ensemble of
identical clumps is, of course, an oversimplification. A realistic
clump size and mass distribution should be taken into account.
Large-scale CO maps present the clumpy structure of the ISM with a
mass distribution following a power law $dN/dM \propto
M^\mathrm{-\alpha}$, where $N$ is the number of clumps and
$\alpha$ has a value around 1.8 \citep{ksrc98b}. Furthermore, the
observations also show that there is a strong correlation of the
density and mass of the clumps corresponding a mass-size relation
$r \propto M^\mathrm{\gamma}$ with $\gamma \approx$ 2.3
\citep{hbsfp98}. For an ensemble of randomly positioned clumps
with a power law mass spectrum, there is a relation among the
power law spectral index $\beta$ of the power spectrum, $\alpha$
and $\gamma$: $\beta = \gamma (3 - \alpha)$ \citep{sbhoz98}. Using
the $\alpha$ and $\gamma$ values above, $\beta=2.76$, which is
close to the power law index $\beta = 2.71$ of $^{13}$CO 2--1 in
IC\,348 found by \citet{skobsm06}.

In this Section we assume these characteristics to be universal
and model the emission by ensemble averaging the PDR single clump
results over such a clump distribution \citep{cub05,croks08} to
reproduce the observed line intensities. The remaining parameter
space is then given by the FUV field input $\chi_{\rm ens}$, the
total mass within the beam $M_\mathrm{ens}$, the average clump
ensemble density $n_\mathrm{mean}$, and the minimum and maximum
clump masses, $m_\mathrm{cl}^\mathrm{min}$ and
$m_\mathrm{cl}^\mathrm{max}$.

To constrain the fit to a three-dimensional parameter space also
for this case, we fix the clump mass limits. The upper mass limit
has to reflect the total mass of the clump distribution because a
spectral index $\alpha=1.8$ corresponds to a top-heavy function,
i.e. a distribution where most of the mass is contained in the few
most massive clumps. We assume that the total mass obtained from
the Ensemble 1 model also provides a reasonable guess for the
total mass in Ensemble 2, consequently setting the upper mass
limit $m_\mathrm{cl}^\mathrm{max}$ to half of the total mass of
Ensemble 1. We set the lower mass limit,
$m_\mathrm{cl}^\mathrm{min}$, to 10$^{-2}$ M$_{\odot}$. These
small clumps can only exist transiently and evaporate on a time
scale of $\sim$ 5000 yrs \citep{kcr08}. The choice of the lower
mass limit is not critical for the fit. \citet{croks08} studied
the dependencies of the model output on the clump mass limits and
found that mainly high - $J$ CO transitions (like $^{12}$CO 8--7)
are sensitive to the lower clump mass limit.

\begin{table*}
\caption{Results of ensembles with a clump mass distribution and
mass-size distribution, {\bf Ensemble case 2}. Column (2) - (4)
present the ratios (in erg\,s$^{-1}$\,sr$^{-1}$\,cm$^{-2}$) of the
modelled and observed line intensities for $^{12}$CO 4--3,
$^{12}$CO 1--0 and $^{13}$CO 1--0; The mean clump ensemble density
$n_\mathrm{mean}$ and the fitted FUV field $\chi_\mathrm{en2}$ are
presented in Column (5) and (6). The C, CO and H$_{2}$ column
densities from the ensembles are presented in Column (7), (8) and
(9); Column (10) is the C/CO ratios. The total masses are listed
in Column (11); Column (12) is the minimum reduced $\chi^{2}$ of
the fit.}
\label{ensemble2}      % is used to refer this table in the text
\centering                          % used for centering table
\begin{tabular}{lccccrrrrccr}        % centered columns (4 columns)
\hline\hline                 % inserts double horizontal lines
Position             & $R_\mathrm{^{12}CO 4-3}$ & $R_\mathrm{^{12}CO 1-0}$ & $R_\mathrm{^{13}CO 1-0}$ & $n_\mathrm{mean}$   &$\chi_\mathrm{en2}$& $N$(C)$_\mathrm{ens}$ & $N$(CO)$_\mathrm{ens}$ & $N$(H$_{2}$)$_\mathrm{ens}$ & C/CO$_\mathrm{ens}$ & $M_\mathrm{ens}$ & $\chi^{2}_{min}$ \\
                     &                          &                          &                          & [10$^{5}$cm$^{-3}$] &                   & 10$^{16}$[cm$^{-2}$]  & 10$^{17}$[cm$^{-2}$]   & 10$^{21}$[cm$^{-2}$]        &                     & M$_{\odot}$      &                  \\
\hline        %inser
Northern rim         & 1.00                     & 1.04                     & 1.44                     & 3.2                &   3                & 14.36                 & 51.82                  & 19.28                       & 0.028               & 3.84             &  1.43            \\
Cluster              & 1.06                     & 1.03                     & 1.13                     & 3.2                & 100                &  9.88                 & 25.58                  &  9.93                       & 0.039               & 1.97             &  0.23            \\
Ridge                & 1.13                     & 1.26                     & 0.91                     & 3.2                &  32                &  9.05                 & 23.43                  &  9.10                       & 0.039               & 1.83             &  0.50            \\
Map center           & 1.08                     & 1.19                     & 1.08                     & 3.2                &  10                & 13.78                 & 40.71                  & 15.39                       & 0.034               & 3.07             &  0.29            \\
$[\ion{C}{i}]$ peak  & 1.13                     & 1.73                     & 0.86                     & 1.0                &  10                & 31.24                 & 40.16                  & 16.61                       & 0.078               & 2.88             &  2.37            \\
Western rim          & 1.83                     & 2.03                     & 1.13                     & 1.0                &  10                & 30.49                 & 23.72                  & 12.61                       & 0.078               & 2.19             &  3.37            \\
Southern rim         & 1.06                     & 0.84                     & 0.66                     & 0.3                &   1                & 13.58                 &  5.24                  &  2.62                       & 0.259               & 0.53             &  4.91            \\
\hline                                   %inserts single line
\end{tabular}
\end{table*}

The actual fit is performed analogously to the ensemble 1 case.
The results are presented in Table~\ref{ensemble2}. We find a
reasonable agreement between the modelled line intensities from
Ensemble 2 and the observed ones. The intensity ratios between
model and observations range between 0.7 and 2.0. The fitted FUV
field $\chi_\mathrm{en2}$ from Ensemble 2 is consistent with the
field $\chi_{\rm ens}$ from Ensemble 1 and that derived from the
FIR continuum in most positions. The maximum deviation occurs at
the northern rim.

The reduced $\chi^{2}$ values are similar for both ensembles at
values less than 5. In spite of the somewhat more complex model,
the fit is not better than for an ensemble of identical clumps
(Ensemble 1).

\begin{figure*} %[h]
\centering \mbox{
\begin{minipage}[b]{8cm}
\includegraphics[width=8cm,angle=0]{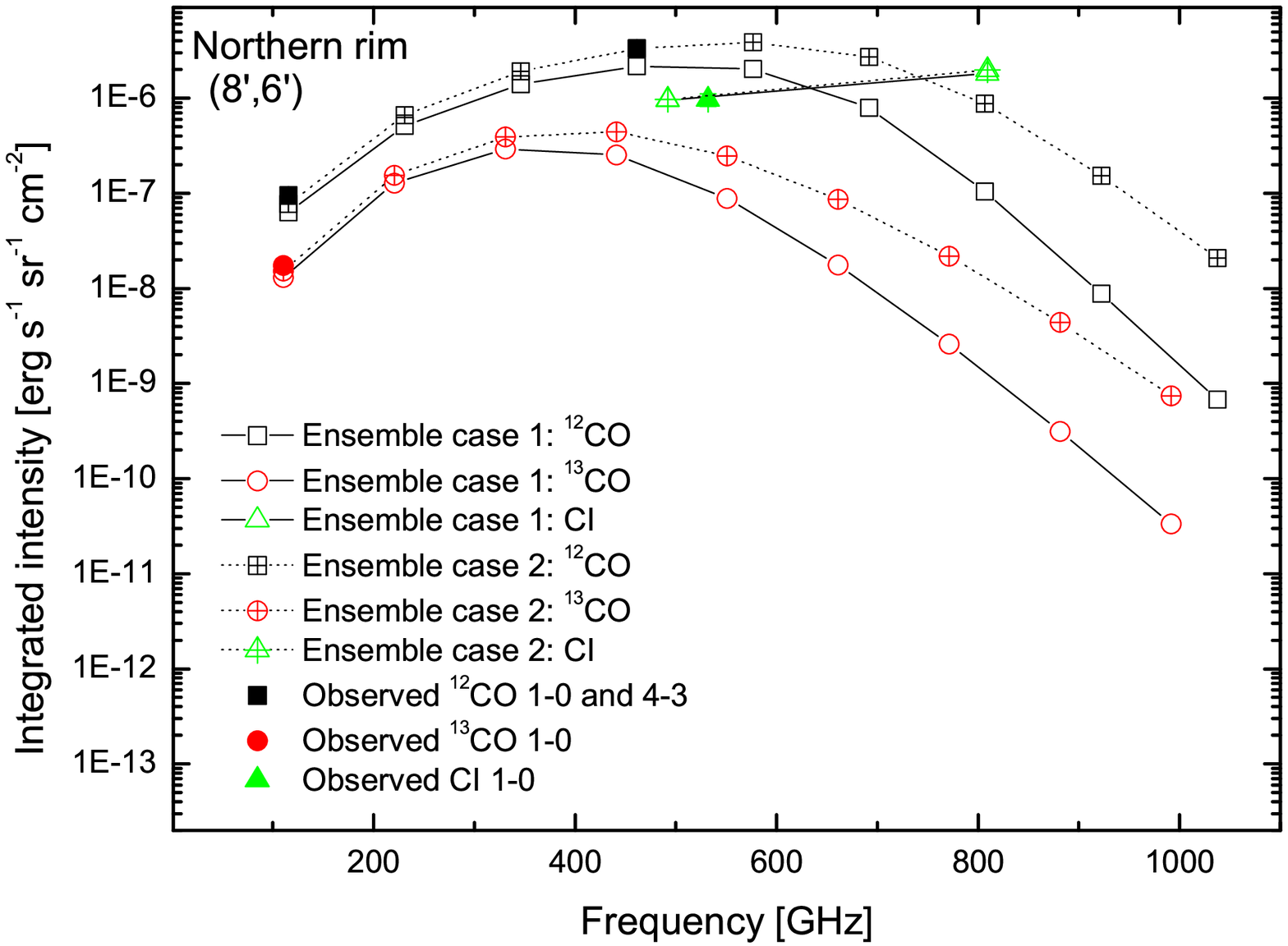}
\end{minipage}
\begin{minipage}[b]{8cm}
\includegraphics[width=8cm,angle=0]{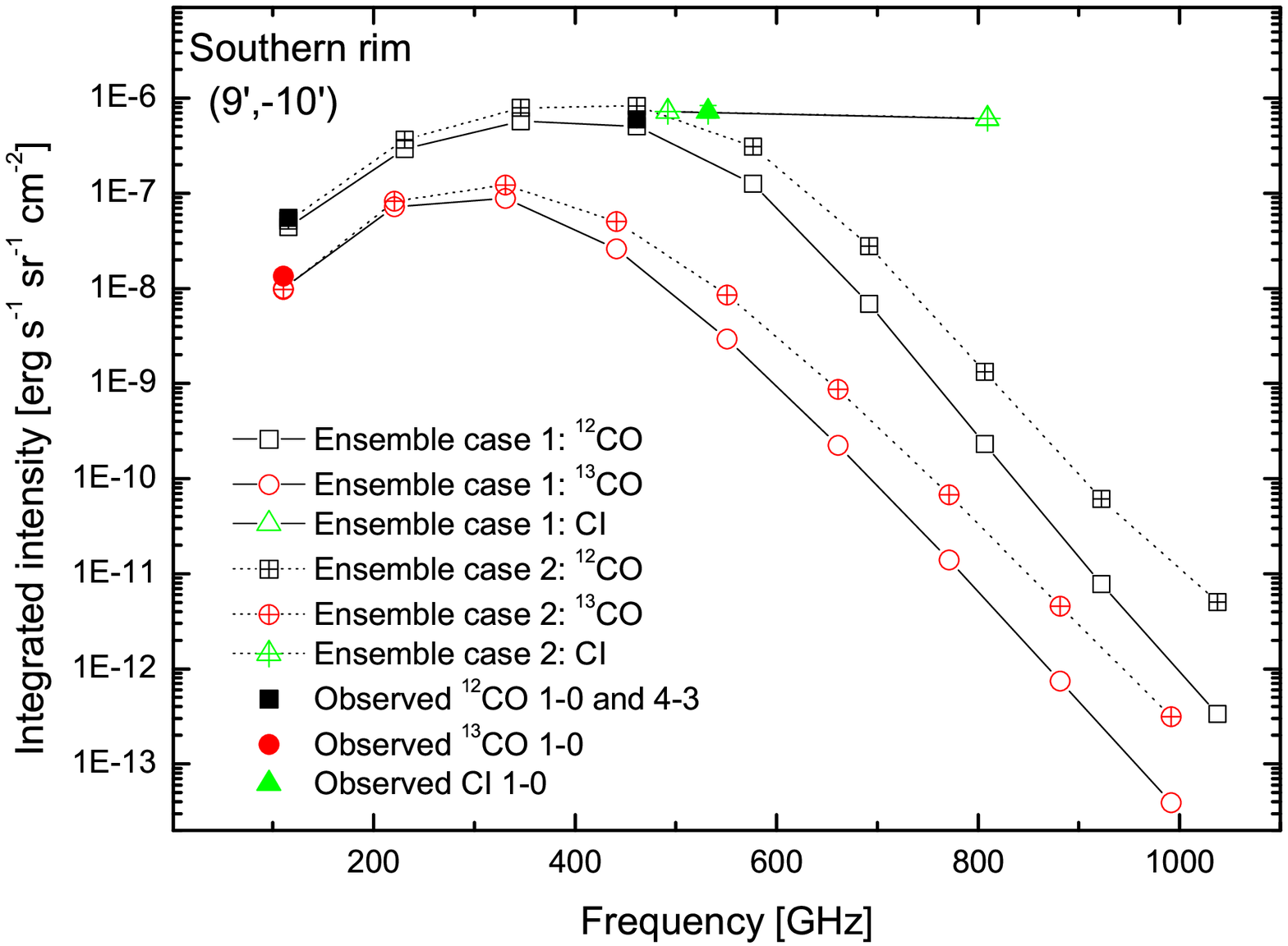}
\end{minipage}
} \caption{Observed (filled symbols) and modelled line intensities
of $^{12}$CO, $^{13}$CO, and [\ion{C}{i}] from both Ensemble case
1 (drawn lines with open symbols) and Ensemble case 2 (dotted
lines, open symbols with a cross inside) at the northern rim (left
panel) and the southern rim (right panel). To better display the
data, we artificially shift the [\ion{C}{i}] frequency by +40\,GHz
for the observed intensity.} \label{int_fre}
\end{figure*}

To compare the observed absolute line intensities with the model
prediction, we show two examples of observed and modelled
integrated intensities versus frequency, namely the cooling curves
of $^{12}$CO, $^{13}$CO (upto $J$ = 9--8) and [\ion{C}{i}] for the
best-fit of both Ensemble 1 and Ensemble 2 models at the northern
rim and the southern rim (see Fig.~\ref{int_fre}). At those two
positions the modelled and observed line intensities agree very
well. Fig.~\ref{int_fre} also provides a prediction for line
intensities of [\ion{C}{i}] $^{3}$P$_{2}$ -- $^{3}$P$_{1}$,
$^{12}$CO 7--6, $^{13}$CO 8--7, etc. Fig.~\ref{int_fre} indicates
that the tracers studied in this paper do not allow to draw any
conclusion about the clump size spectrum that best fits the
observations. Both Ensemble 1 and 2 fit the observed data equally
well. On the other hand, they give very different predictions for
the higher $J$ CO lines. The line intensities predicted for
Ensemble 2 are noticeably larger than those predicted for Ensemble
1 (see Fig.~\ref{int_fre}). The $^{13}$CO 8--7 line, tracing the
column densities of the warm gas, is predicted to be very
sensitive to the differences between the two models, allowing to
discriminate between them.

Based on the CO and H$_{2}$ column densities from the PDR
analysis, we find the averaged CO relative abundance
[CO]/[H$_{2}$] is $\sim$ 2.4\,10$^{-4}$, which is 3 times larger
than the canonical ratio [CO]/[H$_{2}$] of 8\,10$^{-5}$
\citep{flw82} that we used in the LTE analysis. The beam averaged
C/CO abundance ratio (Table~\ref{ensemble2}) varies by a factor 10
between 0.03 at the northern rim, which shows the highest column
density of all studied positions of $2\,10^{21}$\,cm$^{-2}$, and
0.26 at the southern rim, which shows the lowest H$_2$ column
density of $3\,10^{21}$\,cm$^{-2}$. The observed anti-correlation
between C/CO and N(H$_2$) and also the absolute values resemble
those found in many other Galactic molecular clouds as compiled
recently by \citet{mkrm06}. They used the KOSMA-$\tau$ model to
interpret data taken in the Cepheus~B star forming region. They
suggested that the emission of high column density peaks is
dominated by massive clumps exhibiting low C/CO ratios, while
positions of low column densities are dominated by smaller clumps,
exhibiting higher C/CO ratios. This scenario is confirmed in
IC\,348 by a more complete analysis using clump mass ensembles
following the canonical mass and size distributions. However, the
scatter of the present data is large. Observations of the higher
lying transitions of [\ion{C}{i}] and CO are needed to further
improve the analysis.

%________________________________________________________________

\section{Summary and Conclusions}\label{sec_sum}

We have presented fully sampled [\ion{C}{i}] and $^{12}$CO 4--3
maps of the IC\,348 molecular cloud, covering a region of
20\arcmin $\times$ 20\arcmin. The observed $^{12}$CO
4--3/$^{12}$CO 1--0 ratios vary between 0.2 and 1.5. High
$^{12}$CO 4--3/$^{12}$CO 1--0 ratios occur near the B\,5 star, at
the cloud center and northern edge of the cloud.

We have estimated the FUV field from the FIR continuum intensities
obtained from the 60 and 100\,$\mu$m HIRES/IRAS images. The FUV
field in the whole observed region ranges between 1 and 100 Draine
units. We also used HD\,281159, the primary source of UV radiation
in this region, to estimate the FUV field for the seven studied
regions and the FUV field varies between 3 to 40 Draine units.

Using a clumpy PDR model based on the spherical KOSMA - $\tau$ PDR
model we fitted the observed line intensities and intensity ratios
at seven selected positions. The observed intensity ratios provide
a strong constraint on the clump densities. They fall between
4.4\,10$^{4}$\,cm$^{-3}$ and 4.3\,10$^{5}$\,cm$^{-3}$. The FUV
field fitted in the model falls between 2 to 100 Draine units,
consistent with independent estimates for the FUV field derived
from the FIR continuum maps by IRAS and from the stellar
radiation. We found that both an ensemble of identical clumps
(Ensemble 1) and a model using a clump size and mass distribution
following the typically observed scaling (Ensemble 2) produce line
intensities which are in good agreement (within a factor $\sim 2$)
with the observed intensities. The models confirm the
anti-correlation between C/CO abundance ratio and hydrogen column
density found in many regions and explained by \citet{mkrm06}.

The results of this study strengthen the picture of molecular
clouds as highly fragmented, surface dominated objects. Although
Ensemble 2 is the natural case in this picture, Ensemble 1 fits
the observed intensities equally well. We predicted line
intensities for [\ion{C}{i}], CO, and $^{13}$CO transitions for
both cases, which can be tested by future observations. The higher
$J$ CO lines are needed to decide on the necessity of the clump
mass and size distributions and to determine the clump mass limits
\citep{croks08}.

%________________________________________________________________
\begin{acknowledgements}

We thank Naomi Ridge for allowing us to use the FCRAO low-$J$ CO
data in IC\,348. We used the NASA/IPAC/IRAS/HIRES data reduction
facilities. This material is based on work supported by the
Deutsche Forschungs Gemeinschaft (DFG) via grant SFB\,494. This
research has made use of NASA Astrophysics Data System.

\end{acknowledgements}

\end{document}